\documentclass{revtex4}
\usepackage{graphics}
\usepackage{amsmath}
\usepackage{SIunits}

\newcommand{\ben}{\begin{equation}}
\newcommand{\een}{\end{equation}}
\newcommand{\bean}{\begin{eqnarray}}
\newcommand{\eean}{\end{eqnarray}}

\newcommand{\be}{\[}
\newcommand{\ee}{\]}
\newcommand{\bea}{\begin{eqnarray*}}
\newcommand{\eea}{\end{eqnarray*}}

\newcommand{\latin}[1]{\textit{#1}}

\renewcommand{\unit}[2]{\ensuremath{#1\qsk\mathrm{#2}}}
\addunit{\byte}{B}
\addunit{\mebi}{Mi}

\begin{document}

\title{A Simple Cellular Automaton Model for Influenza A Viral Infections}
\author{Catherine Beauchemin}
\affiliation{Department of Physics, University of Alberta, Edmonton, AB, T6G 2J1, Canada\footnote{Author to whom correspondence should be addressed. Electronic address: \texttt{cbeau@phys.ualberta.ca}}}
\author{John Samuel}
\affiliation{Faculty of Pharmacy and Pharmaceutical Sciences, University of Alberta, Edmonton, AB, T6G 2N8, Canada}
\author{Jack Tuszynski}
\affiliation{Department of Physics, University of Alberta, Edmonton, AB, T6G 2J1, Canada}
\date{\today}

\begin{abstract}
Viral kinetics have been extensively studied in the past through the use of spatially homogeneous ordinary differential equations describing the time evolution of the diseased state. However, spatial characteristics such as localized populations of dead cells might adversely affect the spread of infection, similar to the manner in which a counter-fire can stop a forest fire from spreading. In order to investigate the influence of spatial heterogeneities on viral spread, a simple 2-D cellular automaton (CA) model of a viral infection has been developed. In this initial phase of the investigation, the CA model is validated against clinical immunological data for uncomplicated influenza A infections. Our results will be shown and discussed.
\end{abstract}

\maketitle

\section{Introduction}
In the past two decades, many approaches have been chosen to model different aspects of the immune system. Differential equation models are perhaps the most common and are typically used to simulate the immunological and epidemiological dynamics of particular diseases to try and identify the critical parameters involved \cite{bocharov94,perelson02}. Genetic algorithms have been applied to the modelling of the evolution of diversity and pattern recognition capability in the immune system \cite{forrest93,hightower95,hightower96}. On the theoretical front, some ingenious work \cite{farmer86,percus93,smith99,smith01} has been done to understand aspects of the immune system from the perspective of optimization problems. Efforts to build an immune system tailored for computer networks \cite{forrest01,smith96} have raised our understanding of the crucial nature of certain immune system mechanisms. Cellular automaton (CA) models have been developed to simulate the immune dynamics of particular diseases \cite{zorzenon01}, to model shape space interactions \cite{zorzenon98} based on the network theory introduced by Jerne \cite{jerne74}, and have been chosen as the tool of choice to implement highly complex general immune system simulators \cite{bernaschi01,celada96,klein00,seiden92} that are to be used to run immune system experiments \latin{in machina}. More recently, a promising stage-structured modelling approach was used to study cytotoxic T lymphocyte response to antigen \cite{chao03}.

Of all of the above, ordinary differential equations are the most commonly used method to model disease dynamics. By using ordinary differential equations --- which describe a system assuming it is homogeneous in space --- to study viral kinetic, one chooses not to take into consideration spatial characteristics that could potentially play a nontrivial role in the development and outcome of a viral infection. Here, the authors introduce a cellular automaton model that will be used in later work to study the influence of spatial heterogeneities on the dynamical evolution of a viral infection. In this first step of the project, the CA model is used to simulate an uncomplicated infection with the influenza A virus in order to verify that it is not only capable of replicating the typical shape of an immune response to an uncomplicated viral infection, but that it can also give quantitatively reasonable results when parameterized for a particular viral infection. In Section \ref{fluA}, we give a brief description of the nature of influenza. In Section \ref{CAmodel}, we describe the structure and evolution rules of the CA model. In Section \ref{Results}, we present results from the CA simulations and compare them against data from the literature. Finally, in Section \ref{Params}, we expand on the biological meaning of the model's parameters, how we have arrived at the values we chose for the model, and examine the model's sensitivity to their values.

\section{About Influenza A}
\label{fluA}
Influenza, in humans, is caused by a virus that attacks mainly the upper respiratory tract, the nose, throat and bronchi and rarely also the lungs. According to the World Health Organization (WHO), the annual influenza epidemics affect from 5\% to 15\% of the population and are thought to result in between three and five million cases of severe illness and between 250 000 and 500 000 deaths every year around the world. Most deaths currently associated with influenza in industrialized countries occur among the elderly over 65 years of age \cite{who_flu}.

Influenza viruses are divided into three groups: A, B, and C. Additionally, influenza viruses are defined by two different protein components, known as antigens, on the surface of the virus. They are spike-like features called haemagglutinin (H) and neuraminidase (N) components. Influenza A has two subtypes which are important for humans: A(H3N2) and A(H1N1), of which the former is currently associated with most deaths. Antibodies to H are strain-specific and neutralize the infectivity of the influenza A virus, while antibodies to N have a less protective effect. Thus, it is suggested in \cite{bocharov94} that the the antigenic properties of influenza A virus be associated with the H determinant.

It is the high mutation capability of the influenza virus that makes it a great public health concern. The genetic makeup of influenza viruses allows for frequent minor genetic mutations to take place. This makes it necessary to constantly monitor the global influenza situation in order to adjust the influenza vaccines' virus composition annually to include the most recent circulating influenza A(H3N2), A(H1N1) and influenza B viruses. Additionally, influenza A viruses, including subtypes from different species, can swap, reassort, and merge genetic material resulting in novel subtypes. Three times in the last century, influenza A viruses have undergone major genetic changes mainly in their H-component, resulting in global pandemics and large tolls in terms of both disease and deaths \cite{who_birdflu}.

\section{The Cellular Automaton Model}
\label{CAmodel}
Our CA model considers two species: epithelial cells, which are the target
of the viral infection, and immune cells which fight the infection. The
virus particles themselves are not explicitly considered, rather the
infection is modelled as spreading directly from one epithelial cell to
another. The CA is run on a two-dimensional square lattice where each site
represents one epithelial cell. Immune cells are mobile, moving from one
lattice site to another, and their population size is not constant. The CA
lattice is therefore like a tissue of immobile cells which is patrolled by
the mobile immune cells. The CA is updated synchronously (i.e.\ all cells
are time-stepped at once rather than one cell at a time). The boundary
conditions for both the epithelial and immune cells are toroidal, i.e.\ an
immune cell moving off one edge of the grid are reintroduced at the
opposite edge and an infectious epithelial cell at one edge of the grid can
infect healthy cells located at the opposite edge. Finally, the
neighbourhood of a lattice site is defined as consisting of the 8 closest
sites.  Infected epithelial cells can only infect their 8 surrounding
neighbours and immune cells can move at a speed of 1 site per time step to
any of the 8 neighbour sites or remain in place. Below are the detailed
evolution rules for each of the cell species.

\begin{description}
\item[Epithelial Cells]
An epithelial cell can be in any of five states: healthy, infected, expressing, infectious, or dead. Transitions between epithelial cell states occur as follows:
\begin{itemize}
\item Epithelial cells of all states become dead when they are older than \texttt{CELL\_LIFESPAN}.
\item A healthy epithelial cell becomes infected with probability $\texttt{INFECT\_RATE} / (\text{8 nearest neighbours})$ for each infectious nearest neighbour.
\item An infected cell becomes expressing, i.e.\ begins expressing the viral peptide, after having been infected for \texttt{EXPRESS\_DELAY}.
\item An expressing cell becomes infectious after having been infected for $\texttt{INFECT\_DELAY} > \texttt{EXPRESS\_DELAY}$.
\item Infected, expressing, and infectious cells become dead after having been infected for \texttt{INFECT\_LIFESPAN}.
\item Expressing and infectious cells become dead when ``recognized'' by an immune cell. See below for the meaning of ``recognition.''
\item A dead cell is revived at a rate $\texttt{DIVISION\_TIME}^{-1} \times \text{\# healthy} / \text{\# dead}$. When revived, a dead cell becomes a healthy cell or an infected cell with probability $\texttt{INFECT\_RATE} / (\text{8 nearest neighbours})$ for each infectious nearest neighbour.
\end{itemize}
A simulation is initialized with each epithelial cell being assigned a random age between 0 and \texttt{CELL\_LIFESPAN} inclusively. All epithelial cells start in the healthy state with the exception of a fraction \texttt{INFECT\_INIT} of the total number of epithelial cells which, chosen at random, are set to the infected state.

\item[Immune Cells]
An immune cell can be in any of two states: virgin or mature. A virgin cell is an immune cell that has no specificity. A mature cell is an immune cell that has either already encountered an infected cell or has been recruited by another mature immune cell.
\begin{itemize}
\item Immune cells move randomly on the CA lattice at a speed of one lattice site per time step.
\item An immune cell is removed if it is older than \texttt{IMM\_LIFESPAN}.
\item An encounter between an immune cell and an expressing or infectious epithelial cell requires the immune cell to be in the same site as the expressing or infectious cell.
\item A virgin immune cell becomes a mature immune cell if the lattice site it is occupying is in the expressing or infectious states.
\item A mature immune cell occupying an expressing or infectious lattice site ``recognizes'' the epithelial cell and causes it to become dead.
\item Each recognition event causes \texttt{RECRUITMENT} mature immune cells to be added at random sites on the CA lattice after a delay of \texttt{RECRUIT\_DELAY}.
\item Virgin immune cells are added at random lattice sites as needed to maintain a minimum density of \texttt{BASE\_IMM\_CELL} virgin immune cells.
\end{itemize}
A simulation is initialized with a density of \texttt{BASE\_IMM\_CELL}
virgin immune cells at random locations on the CA lattice, each with a
random age.
\end{description}

Implicit in this model are the following physiological assumptions:
\begin{itemize}
\item Only healthy epithelial cells are able to divide.
\item Immune cells cannot be infected and so only ever die of old age.
\item No memory immune cells are considered. Memory could be added by setting an extended lifespan to a given proportion of immune cells created during a viral invasion.
\item We consider a single viral strain with no mutation.  This
implies the existence of a single, unique, epitope identifying the
infection.
\end{itemize}

\section{Results}
\label{Results}
A great deal is known about the influenza virus including viral structure and composition, the replication process, and even some dynamical data regarding the viral and antibody titer over the course of the infection \cite{kilbourne,fritz99,belz02}. However, key dynamical information such as the flow rate of immune cells within an infected tertiary lymphoid organ, the clearance rate of viral particles, the lifespan of an infected epithelial cell, is either uncorroborated, unknown, or known with poor precision. In our model, all but two parameters have been taken directly or adapted from \cite{bocharov94}. They are presented in Table \ref{params} along with their value and description.%
\begin{table}
\begin{center}
\begin{tabular}{ccl}
\hline\hline
Parameter & Value & \hfill Description\hfill{} \\
\hline
\texttt{grid\_width} & $440$ & Width of the grid (cells) \\
\texttt{grid\_height} & $280$ & Height of the grid (cells) \\
\texttt{FLOW\_RATE} & \unit{6}{{ts}/\hour} & Speed of immune cells (time steps/hour) \\
\texttt{IMM\_LIFESPAN} & \unit{168}{\hour} & Lifespan of an immune cell \\
\texttt{CELL\_LIFESPAN} & \unit{380}{\hour} & Lifespan of a healthy epithelial cell \\
\texttt{INFECT\_LIFESPAN} & \unit{24}{\hour} & Lifespan of an infected epithelial cell \\
\texttt{INFECT\_INIT} & $0.01$ & Fraction of initially infected cells \\
\texttt{INFECT\_RATE} & \unit{2}{\hour^{-1}} & Rate of infection of neighbours \\
\texttt{INFECT\_DELAY} & \unit{6}{\hour} & Delay from infected to infectious \\
\texttt{EXPRESS\_DELAY} & \unit{4}{\hour} & Delay from infected to viral expression \\
\texttt{DIVISION\_TIME} & \unit{12}{\hour} & Duration of an epithelial cell division \\
\texttt{BASE\_IMM\_CELL} & $1.5\times10^{-4}$ & Minimum density of immune cells per epithelial cell \\
\texttt{RECRUIT\_DELAY} & \unit{7}{\hour} & Delay between the recruitment call and the addition of immune cells \\
\texttt{RECRUITMENT} & $0.25$ & Number of immune cells recruited when one recognizes the virus \\
\hline\hline
\end{tabular}
\end{center}
\caption{The model's default parameters extracted from \cite{bocharov94}.}
\label{params}
\end{table}

Due to the limited dynamical information available for influenza A, there are only a few quantitative characteristics against which we can compare our model. They are:
\begin{enumerate}
\item The infection should peak on day 2 (\unit{48}{\hour}) \cite{bocharov94,fritz99}.
\item Over the course of the infection, the fraction of epithelial cells that are dead should be as follows \cite{bocharov94}:
\begin{enumerate}
\item 10\% on day 1 (\unit{24}{\hour});
\item 40\% on day 2 (\unit{48}{\hour});
\item 10\% on day 5 (\unit{120}{\hour}).
\end{enumerate}
\item From \cite{bocharov94}, virus concentration should decline to inoculation level on day 6 (\unit{144}{\hour}).  From \cite{fritz99}, experimental data recovered from 8 volunteers indicated that virus shedding persisted for \unit{5\pm2}{\dday} ($72$--\unit{168}{\hour}).
\item The number of immune cells should peak anywhere between day 2 (macrophages' peak) and day 7 (cytotoxic T cells' and B cells' peak) (\unit{48}{\hour}--\unit{168}{\hour}) \cite{bocharov94}.
\item At their peak, the number of B cells, helper T cells, and cytotoxic T cells should be $100$-fold greater than their normal concentration, while that of plasma cells should be $10^4$-fold greater \cite{bocharov94}.
\end{enumerate}

\begin{figure}
\begin{center}
\resizebox{0.5\textwidth}{!}{\includegraphics{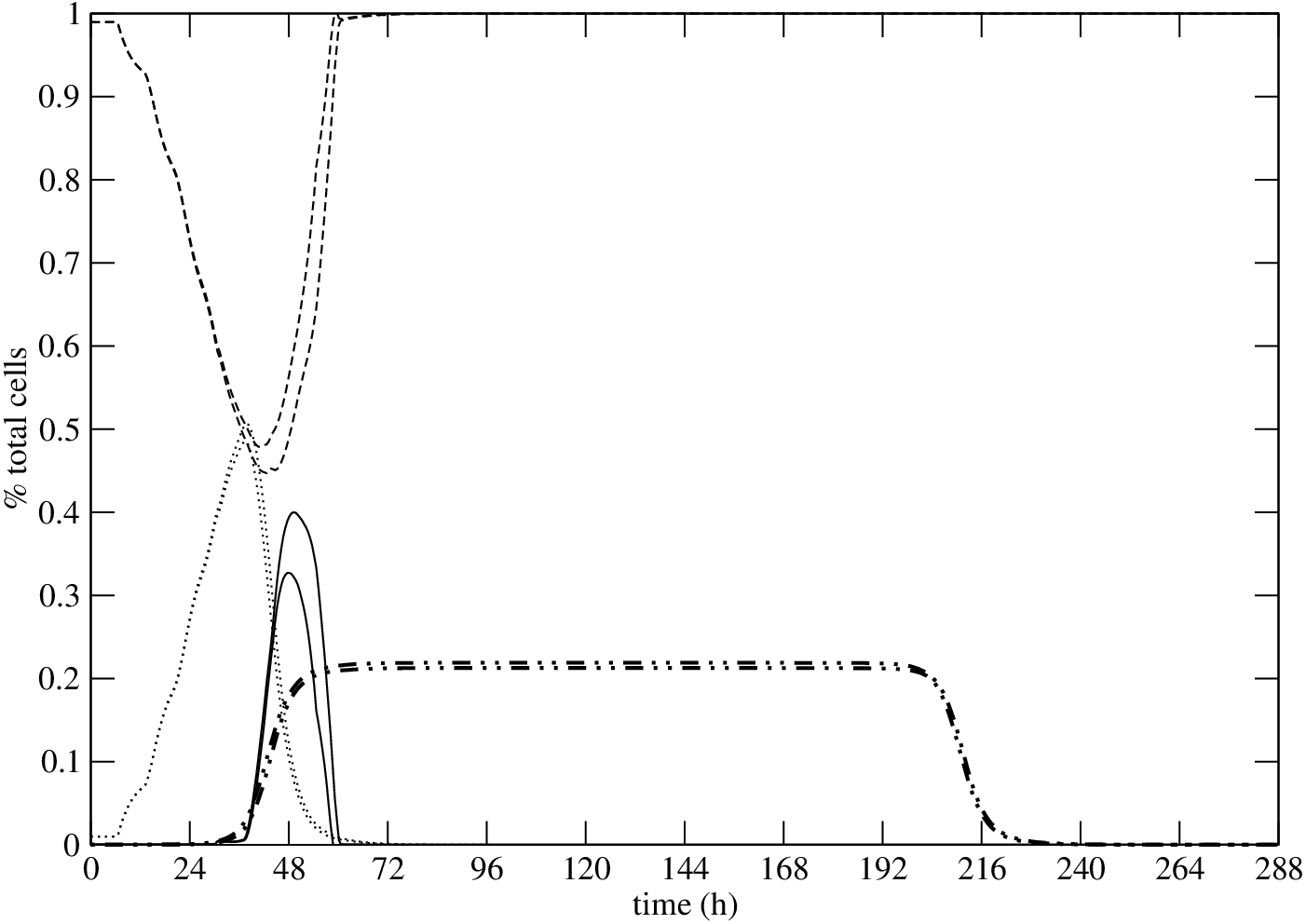}}
\end{center}
\caption{Simulation results averaged over 10 simulation runs using the parameter set presented in Table \ref{params}. The paired lines mark one standard deviation and represent the proportion of epithelial cells that are healthy (dashes), containing the influenza A virus particles (dots), dead (solid), as well as the proportion of immune cells per epithelial cell (dots and dashes).}
\label{simulation}
\end{figure}

The simulation results for the default parameters presented in Table \ref{params} are shown in Figure \ref{simulation}.
The infection peaks on day 2, and although there are fewer than 10\% of the cells dead on days 1 and 5, there is 40\% of the cells dead on day 2. Recovery from the infection is faster than experimental data suggests, but this is probably mainly due to the fact that immune cells in our model are generic. During a true influenza A infection, the first part of the immune response is dominated by macrophages and T cells while the later part is dominated by the action of antibodies, a subtelty that is missed by our bulk model. The number of immune cell does peak between day 2 and 7, with an approximate 1000-fold increase from the normal concentration set by \texttt{BASE\_IMM\_CELL}.

\section{Discussion of the Parameters of the Model}
\label{Params}
Some of the model's parameters have quite a large uncertainty.  In some cases this is a reflection of the difficulty of measuring the parameter while in other cases it is a reflection of the variation between individuals. The model also has two free parameters, \texttt{RECRUITMENT} and \texttt{INFECT\_RATE}, whose values are not known from physiological data. Here, we will discuss each parameter in detail and investigate the sensitivity of the model their variation.

\subsection{\texttt{grid\_width} and \texttt{grid\_height}}
Since each lattice site represents an epithelial cell and the width and height of the grid are given in numbers of epithelial cells, it is possible to calculate the physical size of the simulation. The model's default lattice has $\mathtt{grid\_width} \times \mathtt{grid\_height} = 123,200$ cell sites. In uninflamed tissue, there are about 2,200 ciliated cells per square millimetre of epithelial area \cite{bocharov94}. Taking ciliated cells --- which account for $60\% \sim 80\%$ of internal bronchial surfaces --- to correspond to each lattice site of the CA model, the real-life size of the default lattice is about $\unit{123200}{cells} / \unit{2200}{cells\cdot\milli\meter^{-2}} = \unit{56}{\milli\meter^2}$. When inflamed, there are about 700 cells per square millimetre of epithelial tissue area, giving the simulation a real-life area of \unit{176}{\milli\meter^2}.

\begin{figure}
\begin{center}
\resizebox{0.4\textwidth}{!}{\includegraphics{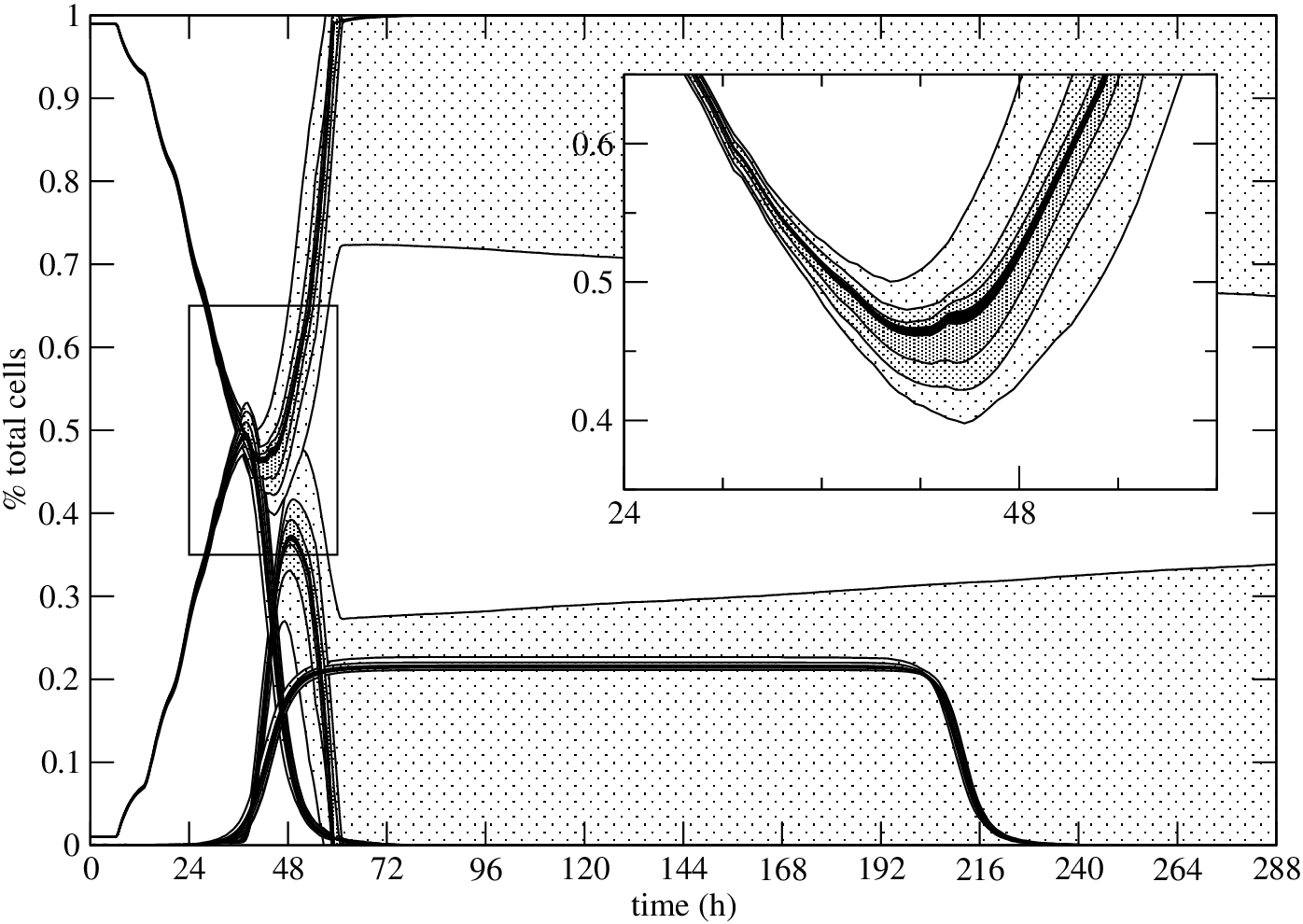}}
\end{center}
\caption{Effects of the lattice size on the dynamics of the infection. The greyed areas represent one standard deviation after 10 runs while increasing darkness represents increasing grid sizes of 7,700 cells, 30,800 cells, 123,200 cells, 1,971,200 cells, and 9,432,500 cells. For clarity, the inset shows a detail of the healthy cell curves alone. Increasing the grid size is seen to not affect the average behaviour but decreases the standard deviation.}
\label{grid_size}
\end{figure}

To test the sensitivity of the model's algorithm to changes in lattice
size, i.e.\ to determine if the finite nature of the simulation influences
the results, simulations have been run with lattices ranging from 7,700
cells to 9,432,500 cells (the most that could be simulated with the
\unit{512}{\mebi\byte} of RAM available in the simulation machine). The
results obtained for several lattice sizes in this range are presented in
Figure \ref{grid_size}.
What one sees from this plot is that the average behaviour of the model is
unchanged by modifications to the size of the simulation area, but that the
standard deviation of the behaviour about the mean decreases as the
simulation area is increased. This is consistent with the interpretation of
the results from a single simulation run on a large grid as being
equivalent to averaged results from several simulations run on smaller
grids. This would be the case --- that a large simulation is equivalent to
an average over several smaller simulations --- anytime the simulation size
itself does not play a role in the dynamics, and so observing this
behaviour strongly suggests that the default lattice size is sufficiently
large for its finite nature to be unimportant.

\subsection{\texttt{FLOW\_RATE}}
Due to the discrete nature of a cellular automaton, immune cells move by increments of one lattice site per time step. Because each lattice site represents one epithelial cell, each time step has to represent the true time required for an immune cell to go from one epithelial cell to the next. Very little is known about the speed at which immune cells patrol tertiary lymphoid organs or by how much the flow will vary depending on whether the area is healthy or infected. This has made the model's time scale difficult to establish.

All time dependent parameters in the model are scaled by a parameter named \texttt{FLOW\_RATE} which is defined as the number of CA time steps corresponding to $\unit{1}{\hour}$ in real time. The duration of the CTL lethal hit process, i.e.\ the time it takes for a cytotoxic T lymphocyte to kill an infected cell, is about $t_{\text{attack}} = \unit{20}{\minute} \sim \unit{30}{\minute}$ \cite{bocharov94}. One can suppose that an immune cell that is not involved with, or in the process of destroying, an infected cell will move from one epithelial cell to the next in less time, say $t_{\text{free}}$. Since on average about 15\% of cells are infected over the course of the infection for the default set of parameters chosen for the model (see Figure \ref{simulation}), one can compute the average time it takes for an immune cell to go from one epithelial cell to the next as $t_{\text{avg}} = 0.15\times t_{\text{attack}}+0.85\times t_{\text{free}}$. Assuming that $t_{\text{free}}$ can be anywhere between negligible, $\approx$ \unit{0}{\minute}, to at most \unit{30}{\minute}, then $t_{\text{avg}}$ can range from $\unit{3}{\minute}$ to $\unit{30}{\minute}$, which corresponds to a range of 2 to 20 time steps per hour for \texttt{FLOW\_RATE}. We have chosen to set \texttt{FLOW\_RATE} to \unit{6}{time\ steps/\hour}, which means that immune cells move at a speed of one epithelial cell per \unit{10}{\minute}. Note that this corresponds to a $t_{\text{free}}$ of about $\unit{6}{\minute} \sim \unit{8}{\minute}$, but we have found no experimental value in the literature to compare this to.

\begin{figure}
\begin{center}
\resizebox{0.3\textwidth}{!}{\includegraphics{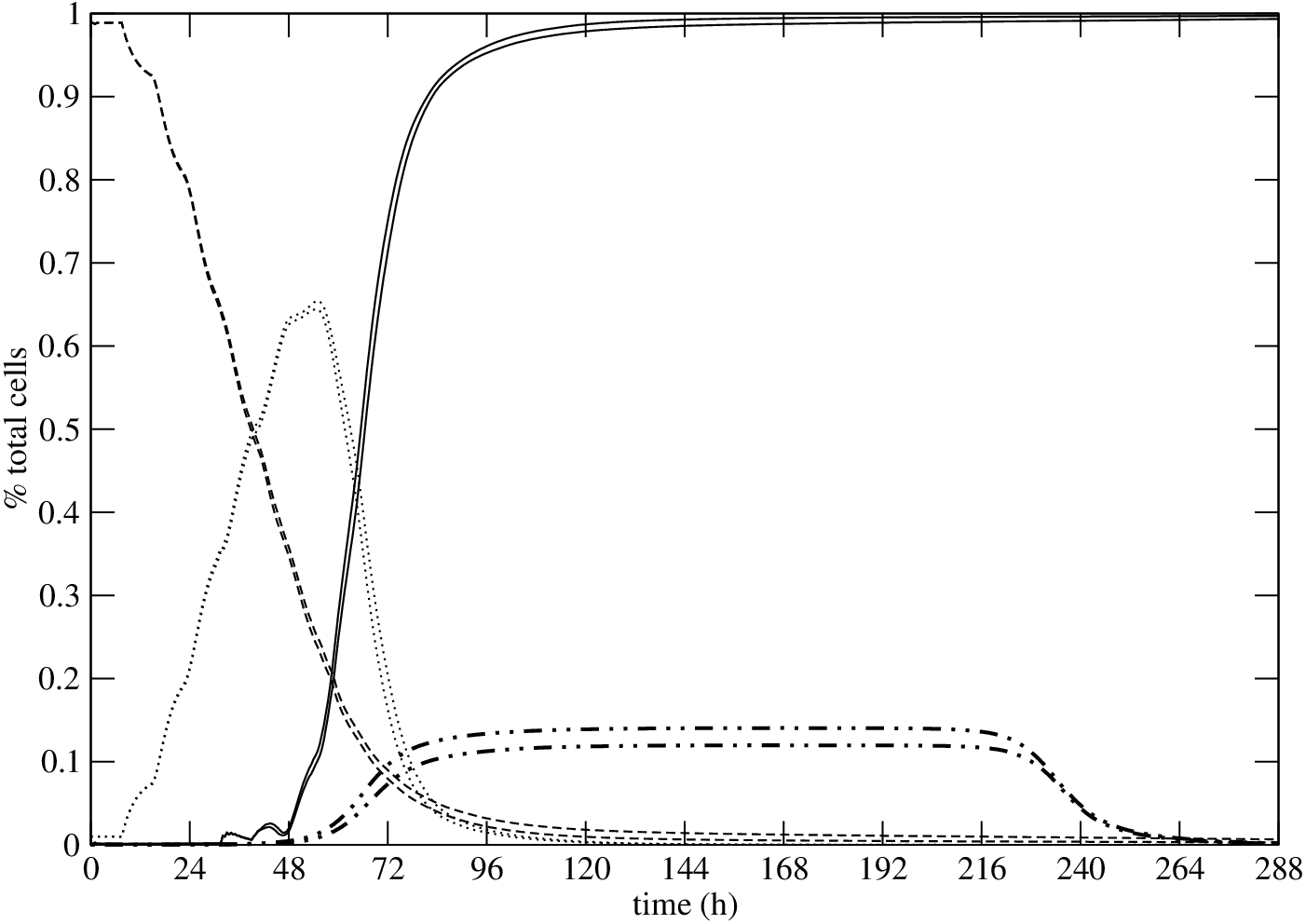}}
\resizebox{0.3\textwidth}{!}{\includegraphics{simulation}}
\resizebox{0.3\textwidth}{!}{\includegraphics{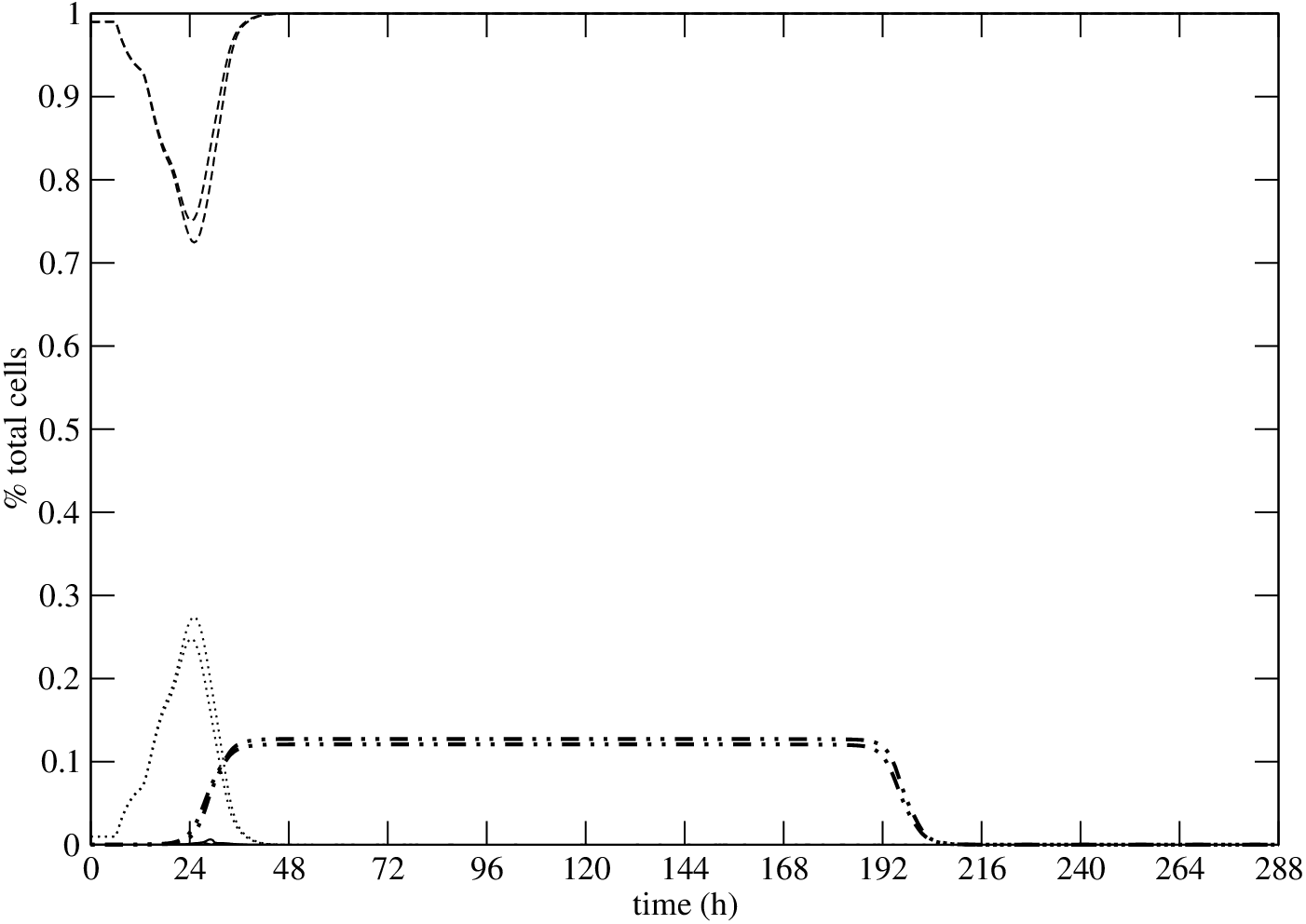}}
\end{center}
\caption{The effect of varying \texttt{FLOW\_RATE} on the viral infection's dynamics. From left to right, the graphs show the behaviour obtained using \texttt{FLOW\_RATE} values of 2, 6, and \unit{20}{time\ steps/\hour} respectively (the physiologically plausible range). The central graph is the same as Figure \ref{simulation}. These show that the simulation is sensitive to the parameter \texttt{FLOW\_RATE}.}
\label{g_flow}
\end{figure}

Figure \ref{g_flow} shows the effect of varying \texttt{FLOW\_RATE} throughout this range.
The faster the immune cells move (the larger \texttt{FLOW\_RATE} is), the bigger the advantage they will have over the growth of the infection and thus the less pronounced the infection. This is clearly illustrated in Figure \ref{g_flow} where the smallest value of \texttt{FLOW\_RATE} results in a lethal infection while the largest value results in a minor infection. The CA model is extremely sensitive to this particular parameter and therefore it would be important for future work to determine this parameter with greater precision.

\subsection{\texttt{IMM\_LIFESPAN}}
\begin{figure}
\begin{center}
\resizebox{0.3\textwidth}{!}{\includegraphics{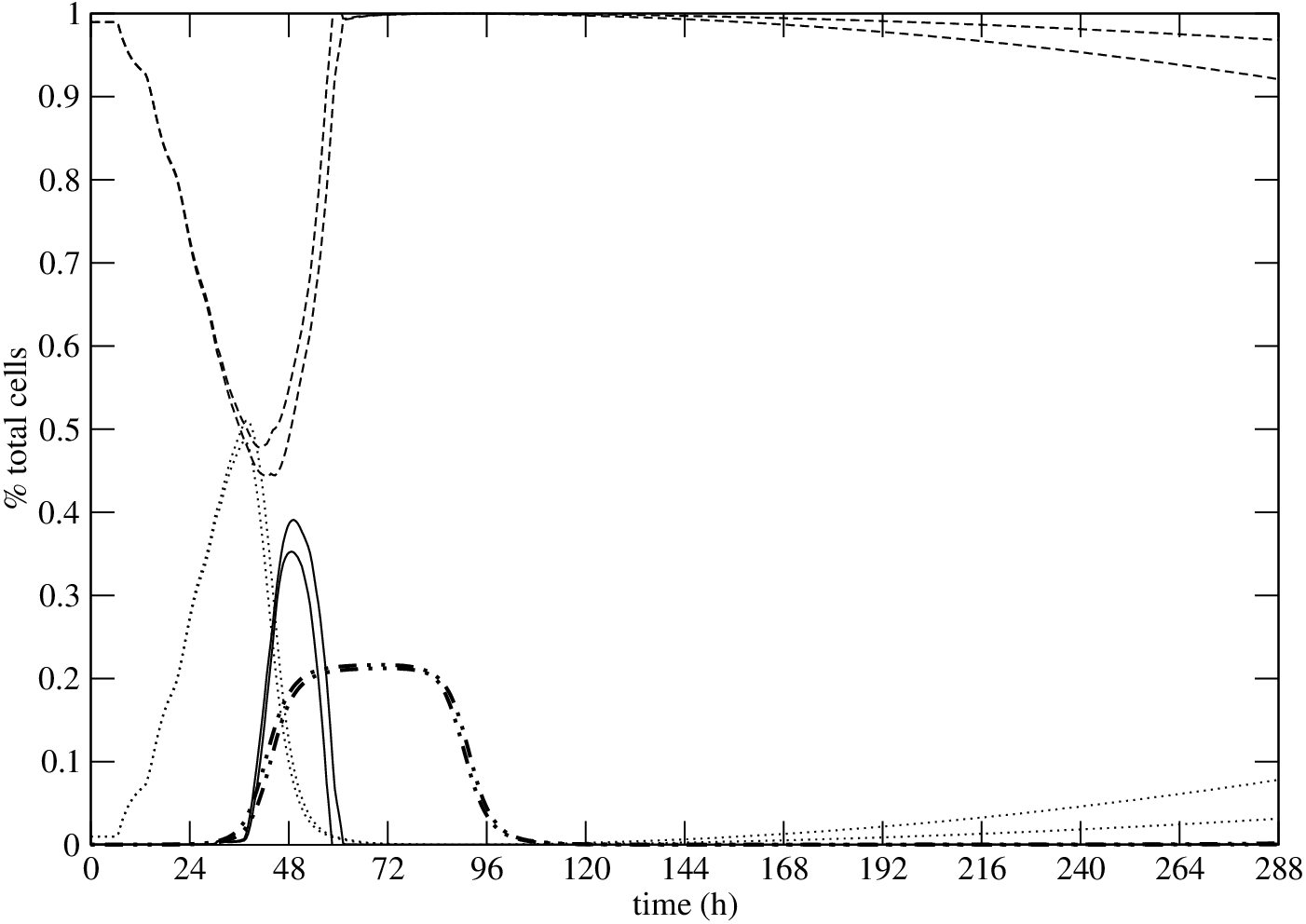}}
\resizebox{0.3\textwidth}{!}{\includegraphics{simulation}}
\resizebox{0.3\textwidth}{!}{\includegraphics{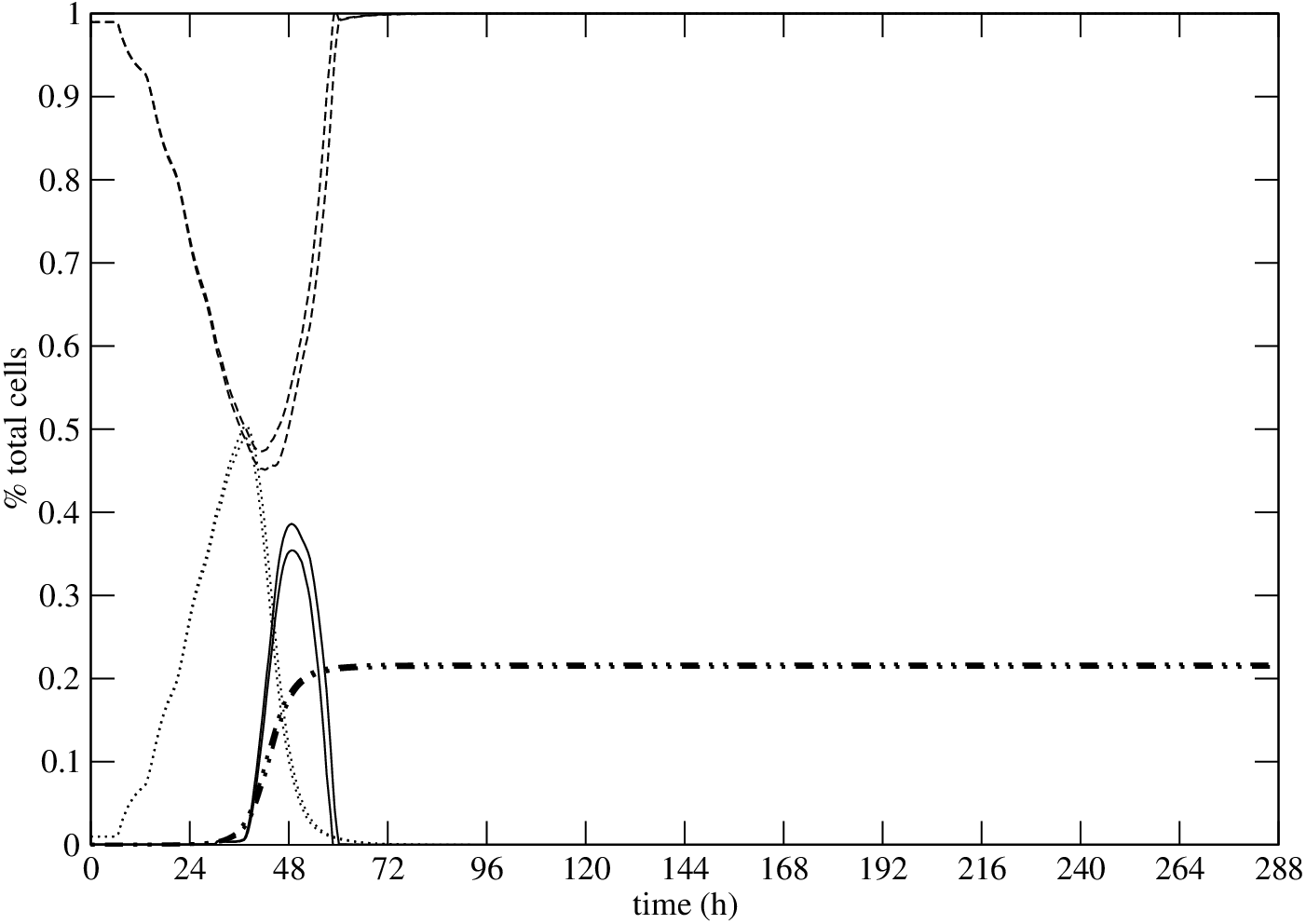}}
\end{center}
\caption{The effect of varying \texttt{IMM\_LIFESPAN} on the viral infection's dynamics. From left to right, the graphs show the behaviour obtained using \texttt{IMM\_LIFESPAN} values of \unit{48}{\hour}, \unit{168}{\hour}, and \unit{480}{\hour} respectively. The central graph is the same as Figure \ref{simulation}. These show that the simulation is only sensitive to the parameter \texttt{IMM\_LIFESPAN} when its value is smaller than the time required for the immune cells to clear the infection.}
\label{g_imm}
\end{figure}

The problem in defining this particular parameter with much precision comes from the fact that the immune cells in our CA model are generic, i.e.\ they do not represent a particular species of immune cell such as cytotoxic T cells or plasma B cells, rather they represent the combined action of all involved species of immune cells. In \cite{bocharov94}, the rate constants of natural death for immune cells range from \unit{0.05}{\dday^{-1}} for B cells to \unit{0.5}{\dday^{-1}} for cytotoxic T cells and plasma B cells. This corresponds to an immune lifespan range of $\unit{48}{\hour} \sim \unit{480}{\hour}$. For our model, we chose an immune cell lifespan of \unit{168}{\hour} as the default value. In Figure \ref{g_imm}, the dynamics resulting from the use of a \unit{48}{\hour} and \unit{480}{\hour} immune cell lifespan is shown.
In broadest terms there are two outcomes and which one is observed depends on whether the immune cell lifespan is greater than or less than the time required for the immune cells to completely eliminate the infection. When the immune cell lifespan is less than the time required to clear the infection, all cell populations oscillate, much in the manner of a predator-prey system --- a situation which may be referred to as a chronic infection. When the immune cell lifespan is greater than the time required to clear the infection, it does just that.

From Figure \ref{g_imm}, it can be seen that only at the very bottom end of the physiologically plausible range for the parameter \texttt{IMM\_LIFESPAN} does its value begin to influence the model's dynamics.  Over most of this parameter's range, the infection is killed too early for the immune cell lifespan to be important. Since experimental data indicates that the virion concentration should have declined to inoculation level no later than \unit{144}{\hour} (\unit{6}{\dday}) into the infection, we chose a value for the immune lifespan that is slightly larger than this, namely \unit{168}{\hour} (\unit{7}{\dday}).

\subsection{\texttt{DIVISION\_TIME}}
The duration of a single division of an epithelial cell is about $\unit{0.3}{\dday} \sim \unit{1}{\dday}$ \cite{bocharov94}. We chose a value of \unit{12}{\hour} (\unit{0.5}{\dday}) for the model parameter \texttt{DIVISION\_TIME}. Assuming that only healthy cells can undergo successful division, the probability per unit time that any dead cell is revived is given by
\be
P(\text{dead} \rightarrow \text{alive})
= \frac{1}{\mathtt{DIVISION\_TIME}} \times \frac{\text{\# healthy cells}}{\text{\# dead cells}}.
\ee

\begin{figure}
\begin{center}
\resizebox{0.3\textwidth}{!}{\includegraphics{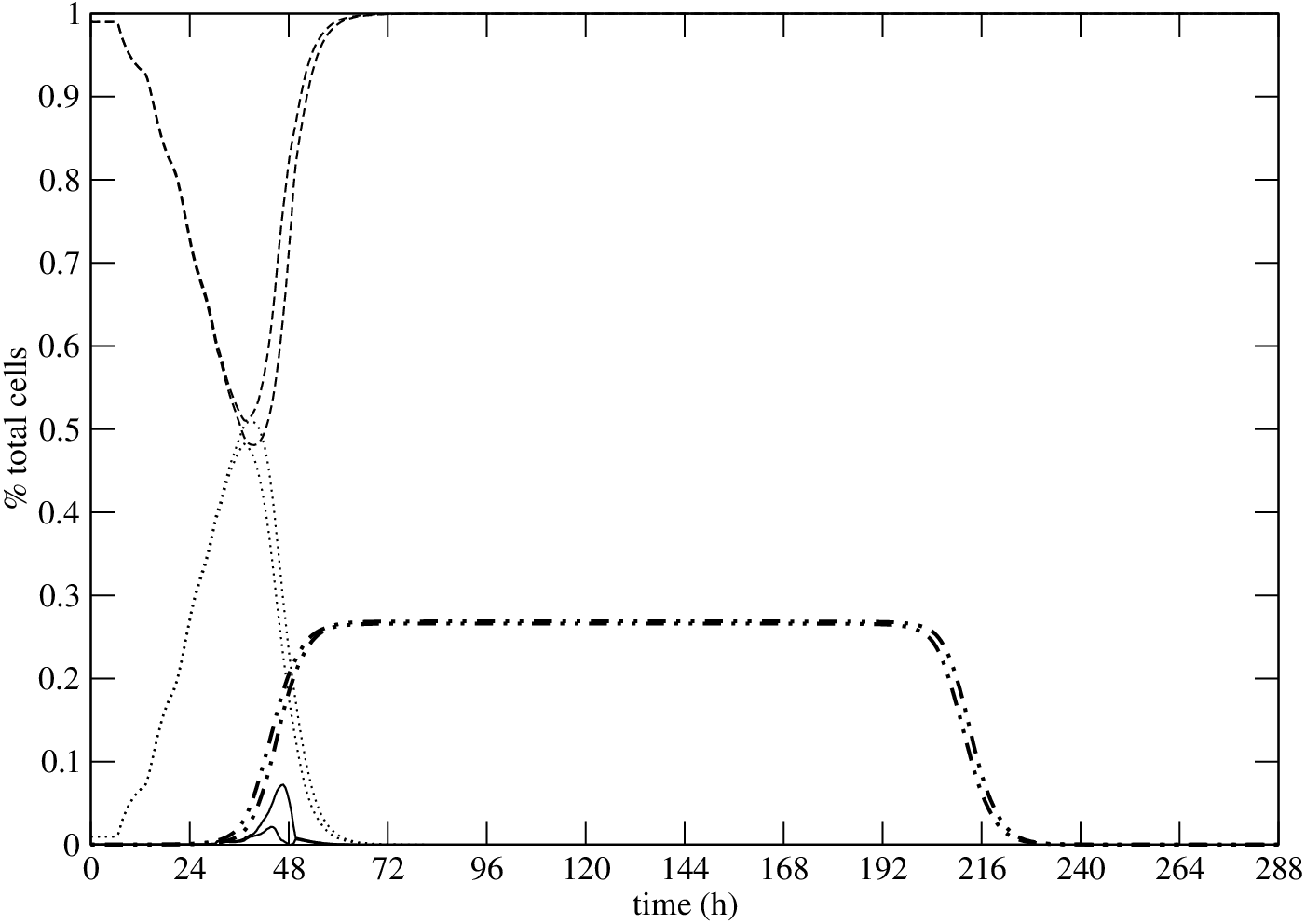}}
\resizebox{0.3\textwidth}{!}{\includegraphics{simulation}}
\resizebox{0.3\textwidth}{!}{\includegraphics{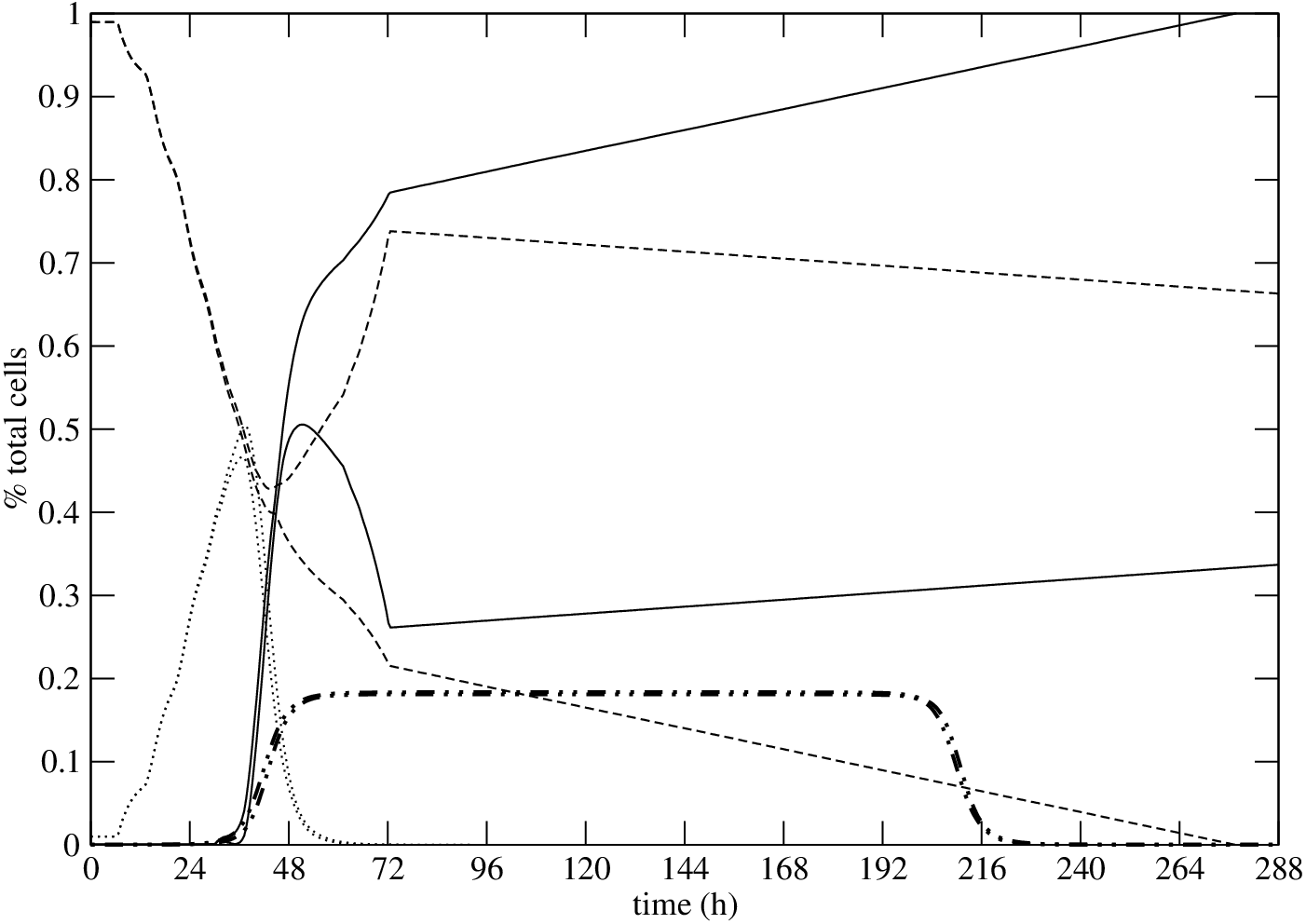}}
\end{center}
\caption{The effect of varying \texttt{DIVISION\_TIME} on the viral infection's dynamics. From left to right, the graphs show the behaviour obtained using \texttt{DIVISION\_TIME} values of \unit{7}{\hour}, \unit{12}{\hour}, and \unit{24}{\hour} respectively. The central graph is the same as Figure \ref{simulation}. These show that the simulation is sensitive to the parameter \texttt{DIVISION\_TIME}.}
\label{g_div}
\end{figure}

Figure \ref{g_div} shows the effect of varying \texttt{DIVISION\_TIME} over its physiologically allowed range of \unit{7}{\hour} to \unit{24}{\hour}.
One sees that this parameter can have a significant effect on the dynamics and even on the outcome of the infection. \texttt{DIVISION\_TIME} represents the capacity of the system to regenerate itself: the greater the regenerative power of the system --- the smaller \texttt{DIVISION\_TIME} --- the better its chances of recuperating after a severe infection. In this model, at the high end of \texttt{DIVISION\_TIME}'s range there is a tendency for the influenza A infection to kill the host, while at the low end of its range the epithelial cell regeneration rate is too high to reproduce the level of damage observed experimentally. We chose a default value of \unit{12}{\hour} for our model. With this value, 40\% of the epithelial cells are dead at the peak of the infection which is in agreement with the experimental data.

\subsection{\texttt{CELL\_LIFESPAN}}
\begin{figure}
\begin{center}
\resizebox{0.3\textwidth}{!}{\includegraphics{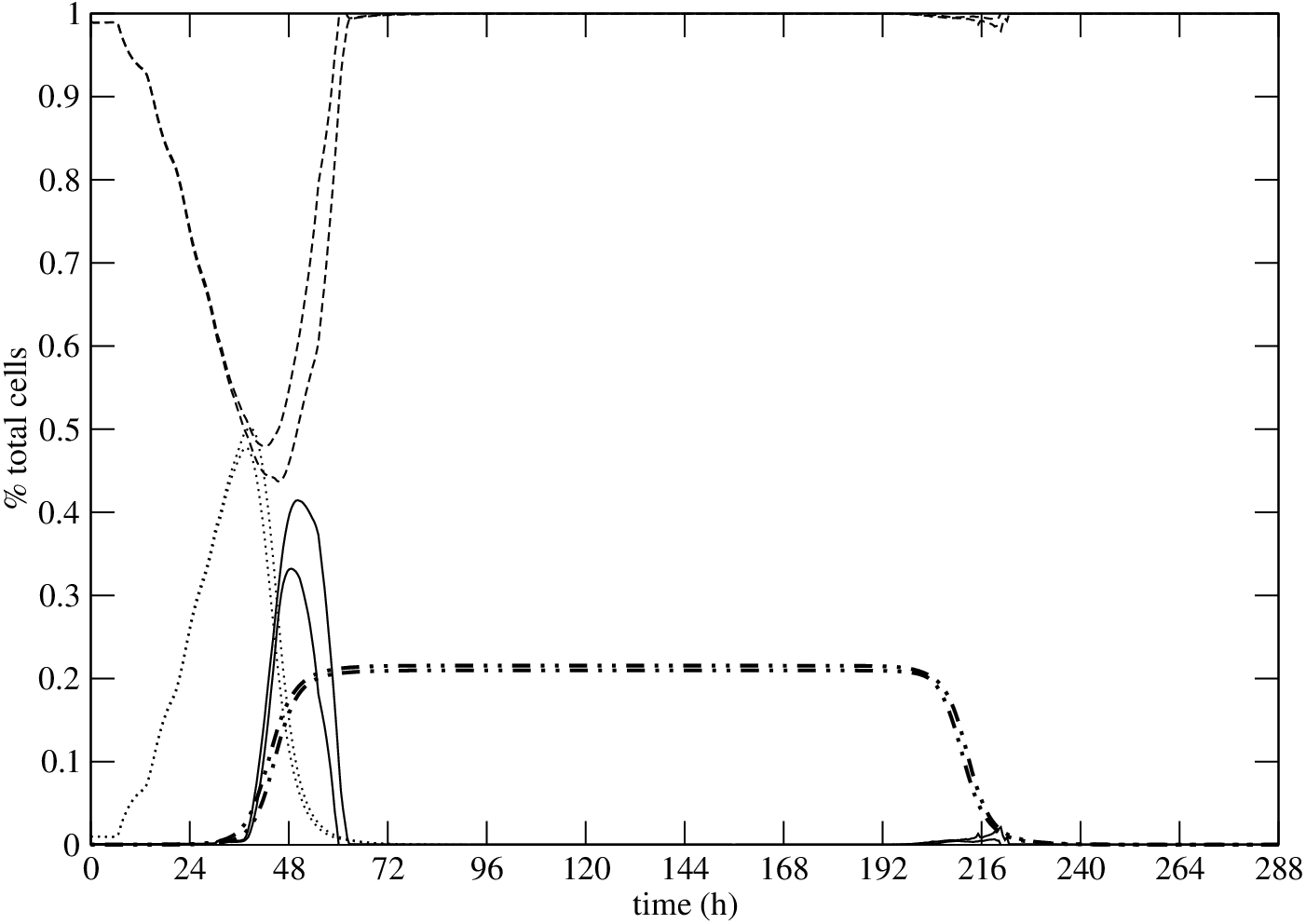}}
\resizebox{0.3\textwidth}{!}{\includegraphics{simulation}}
\resizebox{0.3\textwidth}{!}{\includegraphics{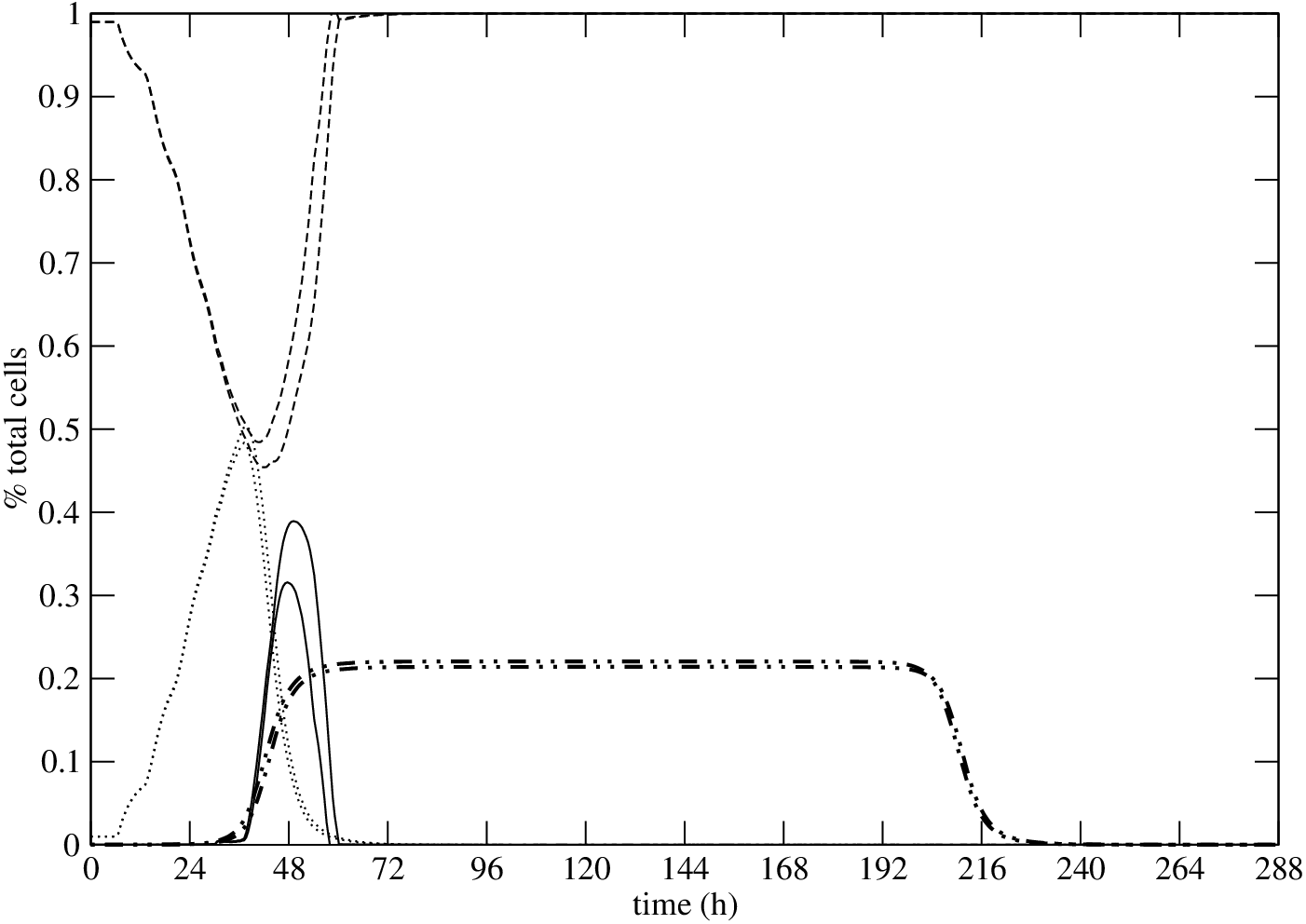}}
\end{center}
\caption{The effect of varying \texttt{CELL\_LIFESPAN} on the viral infection's dynamics. From left to right, the graphs show the behaviour obtained using \texttt{CELL\_LIFESPAN} values of \unit{160}{\hour}, \unit{380}{\hour}, and \unit{600}{\hour} respectively. The central graph is the same as Figure \ref{simulation}. These show that the simulation is not sensitive to the parameter \texttt{CELL\_LIFESPAN}.}
\label{g_cell}
\end{figure}

An epithelial cell, \latin{in vitro}, can go through 20 to 25 population doublings in its lifespan \cite{piao01}.  Each population doubling lasts for \texttt{DIVISION\_TIME}, so the lifespan of an epithelial cell should be about $\unit{160}{\hour} \sim \unit{600}{\hour}$. We chose the mid-range value of \unit{380}{\hour} as the default for our model. One impediment to finding more precise information about epithelial cell lifespans is that most \latin{in vitro} research is done using cell lines that have been made immortal. Additionally, there is much variability in cell lifespan across cell types. In any case, even the lower bound of \unit{160}{\hour} is longer than the duration of an uncomplicated influenza A infection and thus the value of \texttt{CELL\_LIFESPAN}, within the physiologically plausible range, should have little effect on the results. Figure \ref{g_cell} shows the dynamics of the viral infection for cell lifespans at the physiologically allowed extrema of $\unit{160}{\hour}$ and $\unit{600}{\hour}$ as well as at the default mid-range value of $\unit{380}{\hour}$.
In the model, one consequence of the en masse killing of epithelial cells
by the viral infection is the synchronization of the ages of newly-grown
epithelial cells. This is seen as a cell-death ``echo'' approximately
$\unit{\texttt{CELL\_LIFESPAN}}{\hour}$ after the peak of the infection.
The only visible difference between simulations run with different values
of \texttt{CELL\_LIFESPAN} is a change in the time at which the cell-death
echo occurs.

\subsection{\texttt{EXPRESS\_DELAY}, \texttt{INFECT\_DELAY}, and \texttt{INFECT\_LIFESPAN}}
When an epithelial cell becomes infected by an influenza A virion, \unit{4}{\hour} (\texttt{EXPRESS\_DELAY}) elapse before it can start expressing viral antigen and thus be recognized by an immune cell \cite{bocharov94}. It will take the infected cell an additional \unit{2}{\hour} ($\texttt{INFECT\_DELAY}=\unit{6}{\hour}$) to begin releasing virus particles and thus become infectious \cite{bocharov94}. Ultimately, influenza A virus being highly cytopathic, the lifespan of an infected epithelial cell is shortened to at most 1 day ($\texttt{INFECT\_LIFESPAN}=\unit{24}{\hour}$) \cite{bocharov94}.

\subsection{\texttt{INFECT\_INIT}}
\begin{figure}
\begin{center}
\resizebox{0.3\textwidth}{!}{\includegraphics{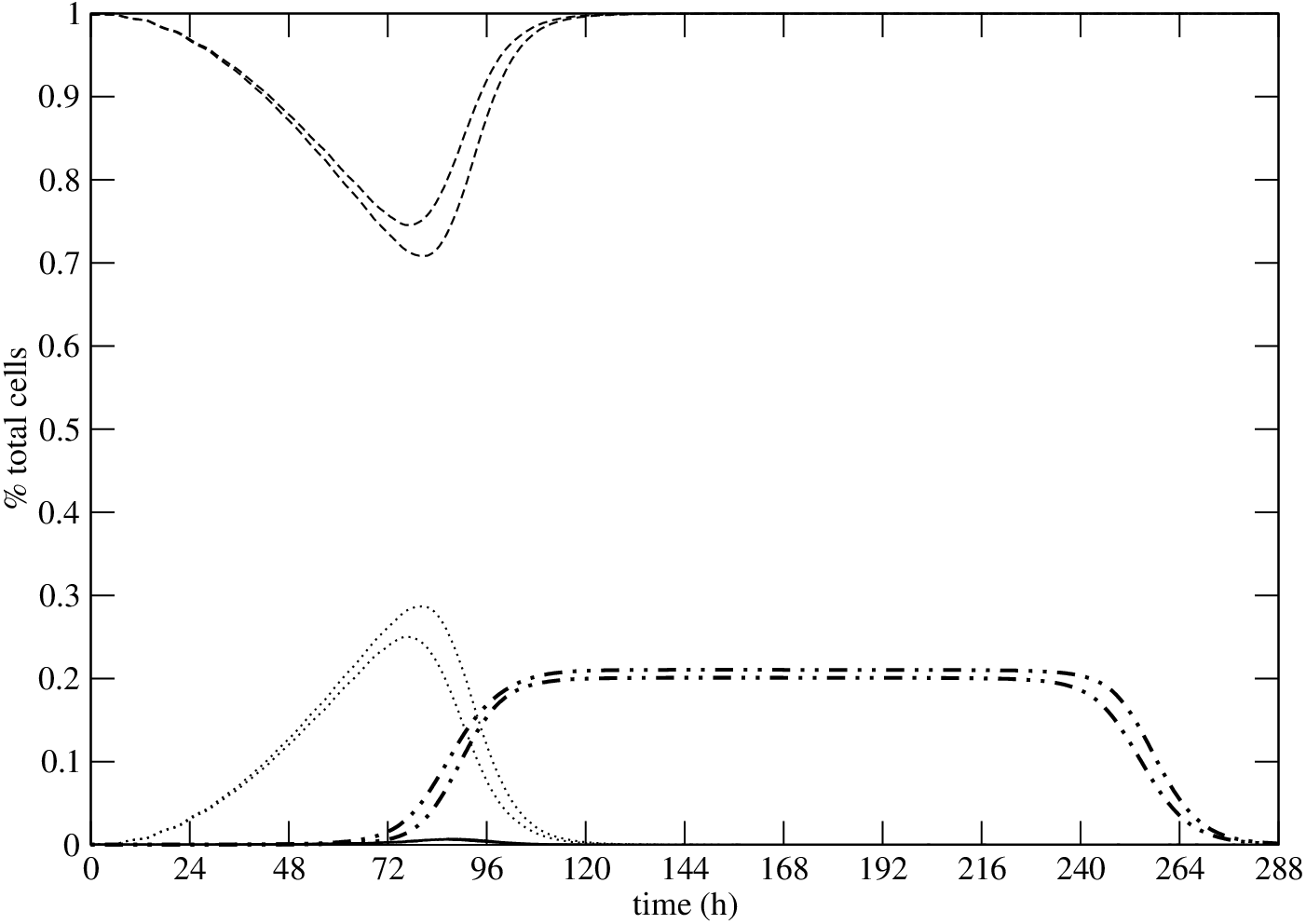}}
\resizebox{0.3\textwidth}{!}{\includegraphics{simulation}}
\resizebox{0.3\textwidth}{!}{\includegraphics{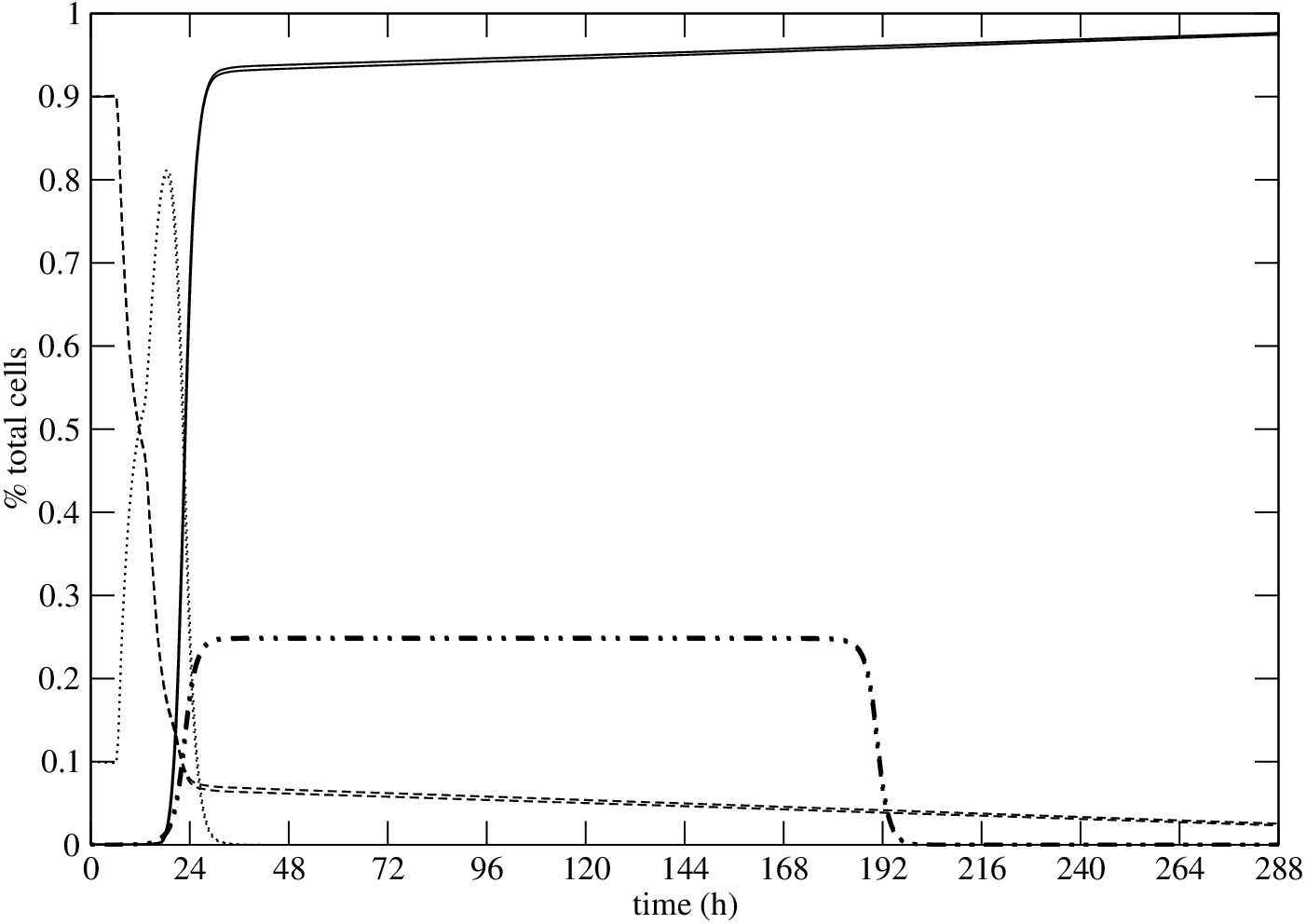}}
\end{center}
\caption{The effect of varying \texttt{INFECT\_INIT} on the viral infection's dynamics. From left to right, the graphs show the behaviour obtained using \texttt{INFECT\_INIT} values of $0.001$, $0.01$, and $0.1$ respectively. The central graph is the same as Figure \ref{simulation}. These show that the simulation is sensitive to the parameter \texttt{INFECT\_INIT}.}
\label{g_init}
\end{figure}

This parameter represents the magnitude of the initial dose of influenza A virus delivered to test volunteers in viral replication experiments.  In particular, in our model this is the fraction of epithelial cells initially set in the infected state. This parameter's value is not bounded physiologically, rather it is an arbitrary initial condition. In studies of the influenza A virus reported in \cite{bocharov94}, the typical initial concentration of aerosol-delivered virions is about \unit{10^8}{virions/\milli\liter}. Since the average concentration of epithelial cells in the area typically affected by influenza varies from $\unit{10^9}{cells/\milli\liter}$ to $\unit{10^{10}}{cells/\milli\liter}$, and a single epithelial cell can absorb from 1 to 10 influenza virions, then the typical fraction of cells initially infected is computed to be
\bea
\mathtt{INFECT\_INIT}
&=& \frac{\unit{10^8}{virions/\milli\liter}}{\unit{(10^9 \sim 10^{10})}{cells/\milli\liter}\times\unit{(1 \sim 10)}{virions/cell}} \\
&=& 0.001 \sim 0.1.
\eea
We have chosen a default value of $0.01$ --- the logarithmic middle of the range. Figure \ref{g_init} presents the effect on the model's dynamics of varying \texttt{INFECT\_INIT} over this range.
One sees that the evolution and the outcome of the infection is dependent upon the initial infective dose received. This reflects the competitive nature of the disease and immune system interaction: if one starts with too large a numerical advantage over the other, the latter is bound to loose.

\subsection{\texttt{INFECT\_RATE}}
\begin{figure}
\begin{center}
\resizebox{0.3\textwidth}{!}{\includegraphics{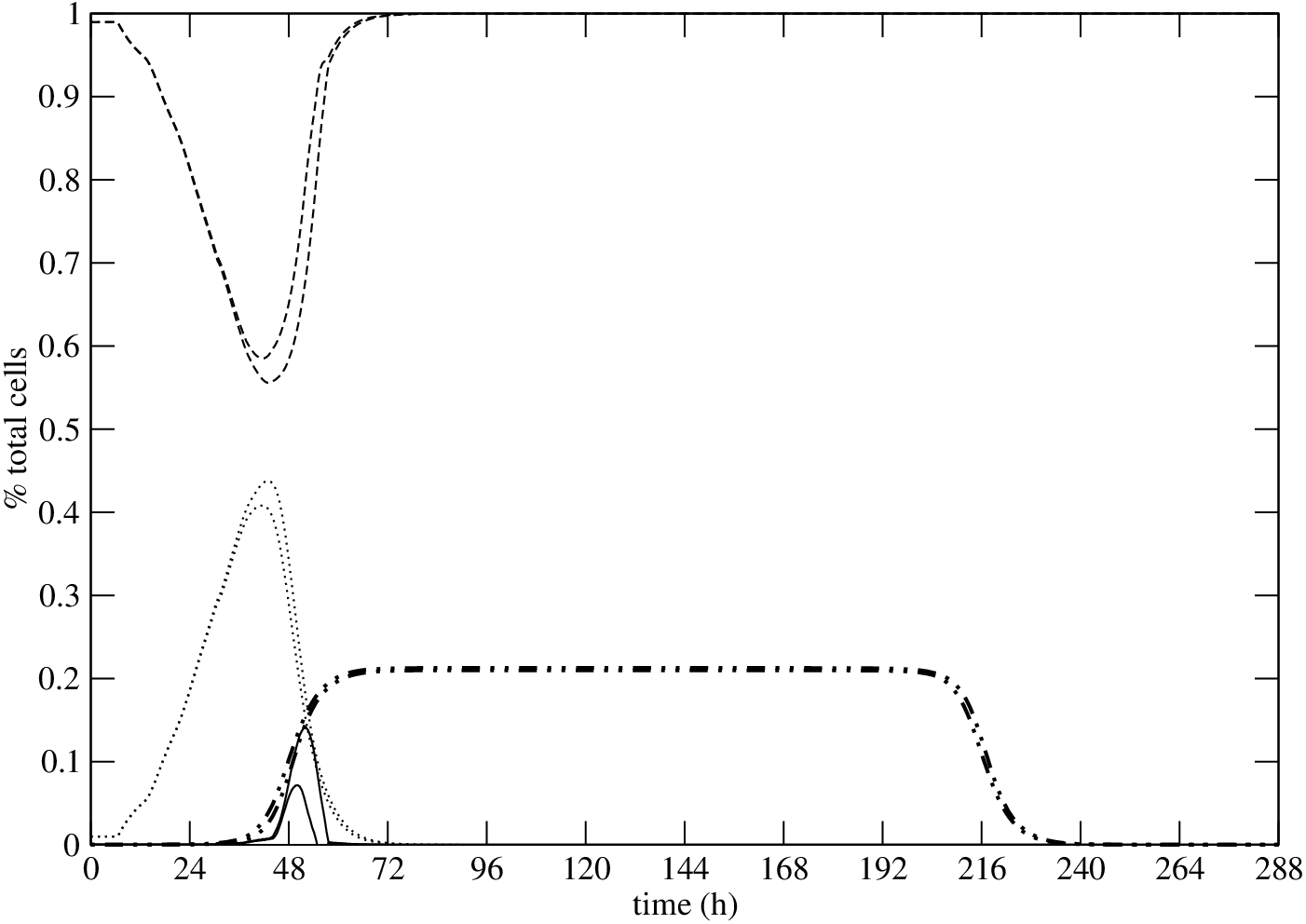}}
\resizebox{0.3\textwidth}{!}{\includegraphics{simulation}}
\resizebox{0.3\textwidth}{!}{\includegraphics{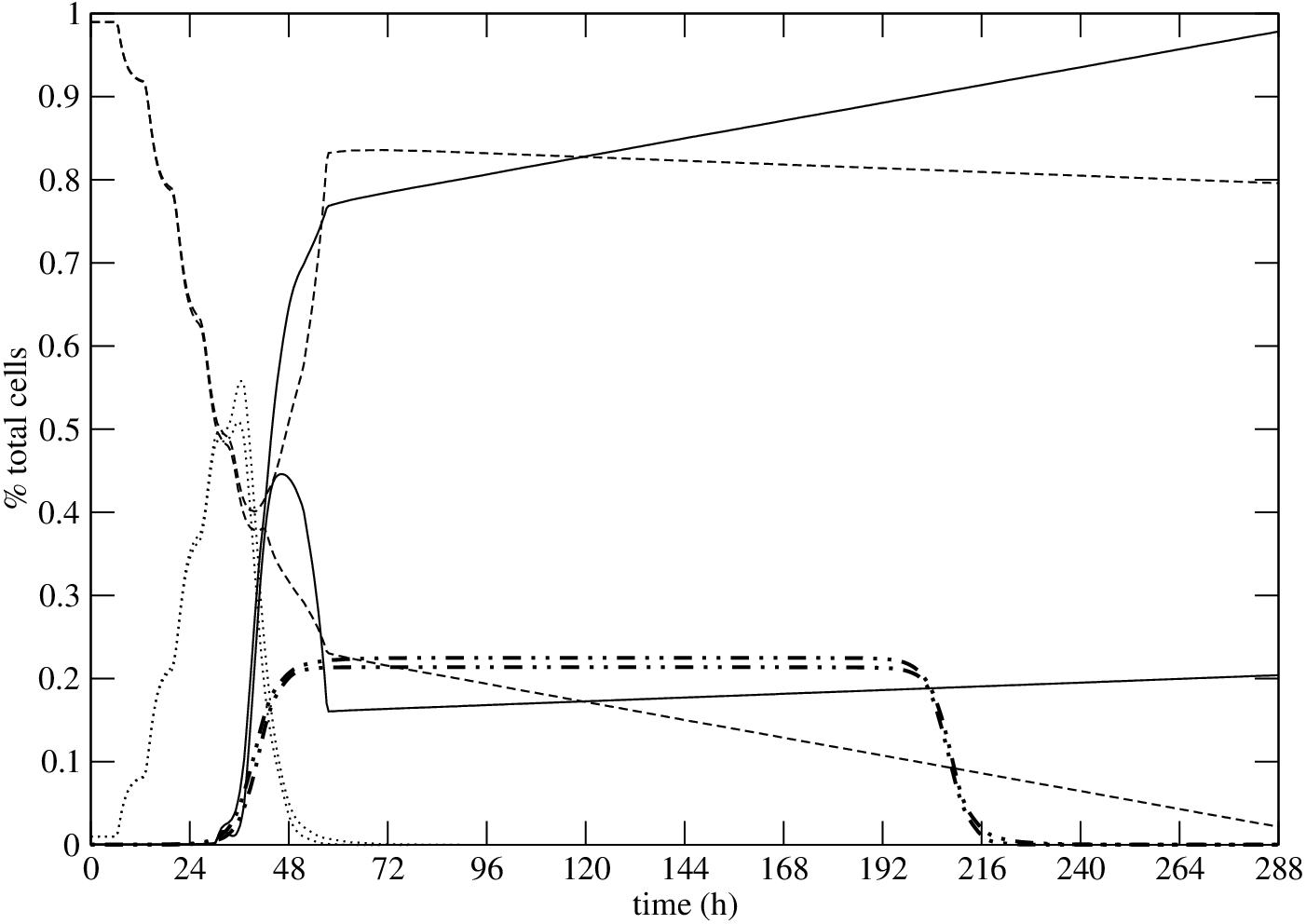}}
\end{center}
\caption{The effect of varying \texttt{INFECT\_RATE} on the viral infection's dynamics. From left to right, the graphs show the behaviour obtained using \texttt{INFECT\_RATE} values of 1, 2, and \unit{4}{infected\ neighbour(s)\cdot\hour^{-1}} respectively. The central graph is the same as Figure \ref{simulation}. These show that the simulation is sensitive to the parameter \texttt{INFECT\_RATE}.}
\label{g_inf}
\end{figure}

The growth rate of the infection in our model is determined by the parameter \texttt{INFECT\_RATE}, which is defined as the number of its neighbours that an infectious cell will infect per hour. This parameter was not found in the literature as it does not correspond directly to the true biological process of infection. In fact, \unit{6}{\hour} after a cell has been infected by the influenza A virus, it begins to release influenza A virions (virus particles) into its surrounding and it is those virions that move on to infect neighbouring cells. In our model, we have decided not to consider virions at all and instead have modelled the infection process by the direct infection of neighbouring cells by infectious cells. This is computationally simpler and more efficient. Throughout influenza A literature, however, the infection growth is described in terms of the viral titer over the course of the infection rather then as a growth rate. This makes it difficult to establish the parameter \texttt{INFECT\_RATE} from experimental data. We have chosen to make it a free parameter and found that we obtained the best results by setting it to \unit{2.0}{\hour^{-1}}, which means that an infectious cell will, on average, infect 2 of its 8 nearest neighbours per hour. One expects that a smaller value of this parameter will yield a less pronounced infection while larger values will increase the amplitude of the infection and potentially result in the death of the host. This is illustrated in Figure \ref{g_inf} where we have explored the effect of using a value $1/2 \times$ (\unit{1}{infected\ neighbour\cdot\hour^{-1}}) and a value $2 \times$ (\unit{4}{infected\ neighbours\cdot\hour^{-1}}) the default.

\subsection{\texttt{BASE\_IMM\_CELL}}
The viral infection model only concerns itself with immune cells that have a role to play in the viral infection, i.e.\ those immune cells with a receptor capable of recognizing an epitope of the viral strain present in the simulation. There must always remain a minimum density of cells capable of recognizing and responding to a new outbreak, and this density is set by the parameter \texttt{BASE\_IMM\_CELL}. The value for \texttt{BASE\_IMM\_CELL} was calculated based on the following information. It has been determined \cite{westermann92} that intraepithelial lymphocytes have a frequency of about 15 per 100 epithelial cells. Further, the probability that an immune cell chosen at random will recognize an epitope chosen at random is typically $10^{-5}$ \cite{perelson97}. Finally, taking into account memory and cross-reactivity phenomena in the case of influenza A in adults, it is suggested in \cite{bocharov94} that the fraction of B or T cells capable of responding to a viral infection with influenza A be increased by 100-fold. Thus, \texttt{BASE\_IMM\_CELL} is calculated to be $(15 / 100) \times 10^{-5} \times 100 = 1.5 \times 10^{-4}$ virgin immune cells per epithelial cell.

\subsection{\texttt{RECRUITMENT}}
When a mature immune cell encounters an infected epithelial cell presenting a viral peptide, the mature immune cell will ``recruit'' a number of new mature immune cells. The number of newly created mature immune cells is set through the parameter \texttt{RECRUITMENT}. This parameter is one of the two free parameters of the model, and it has no direct biological basis because the immune cells in our model are generic. This parameter represents the efficiency with which the immune system responds to the viral infection. We explored a range of values for \texttt{RECRUITMENT}, and found that a value of $0.25$, i.e.\ a 25\% chance of recruiting a mature immune cell upon each successful recognition, gives satisfying results.

\begin{figure}
\begin{center}
\resizebox{0.3\textwidth}{!}{\includegraphics{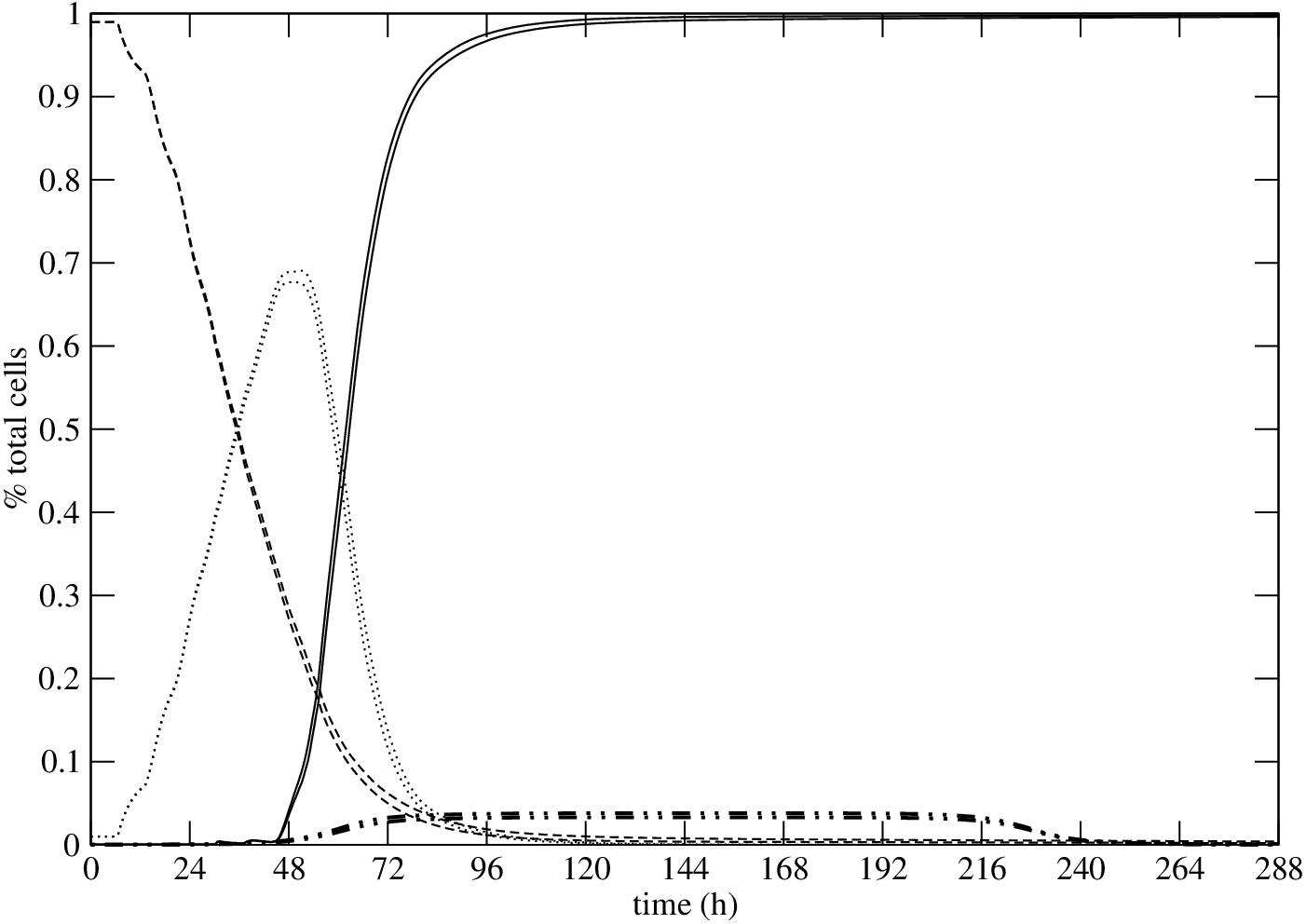}}
\resizebox{0.3\textwidth}{!}{\includegraphics{simulation}}
\resizebox{0.3\textwidth}{!}{\includegraphics{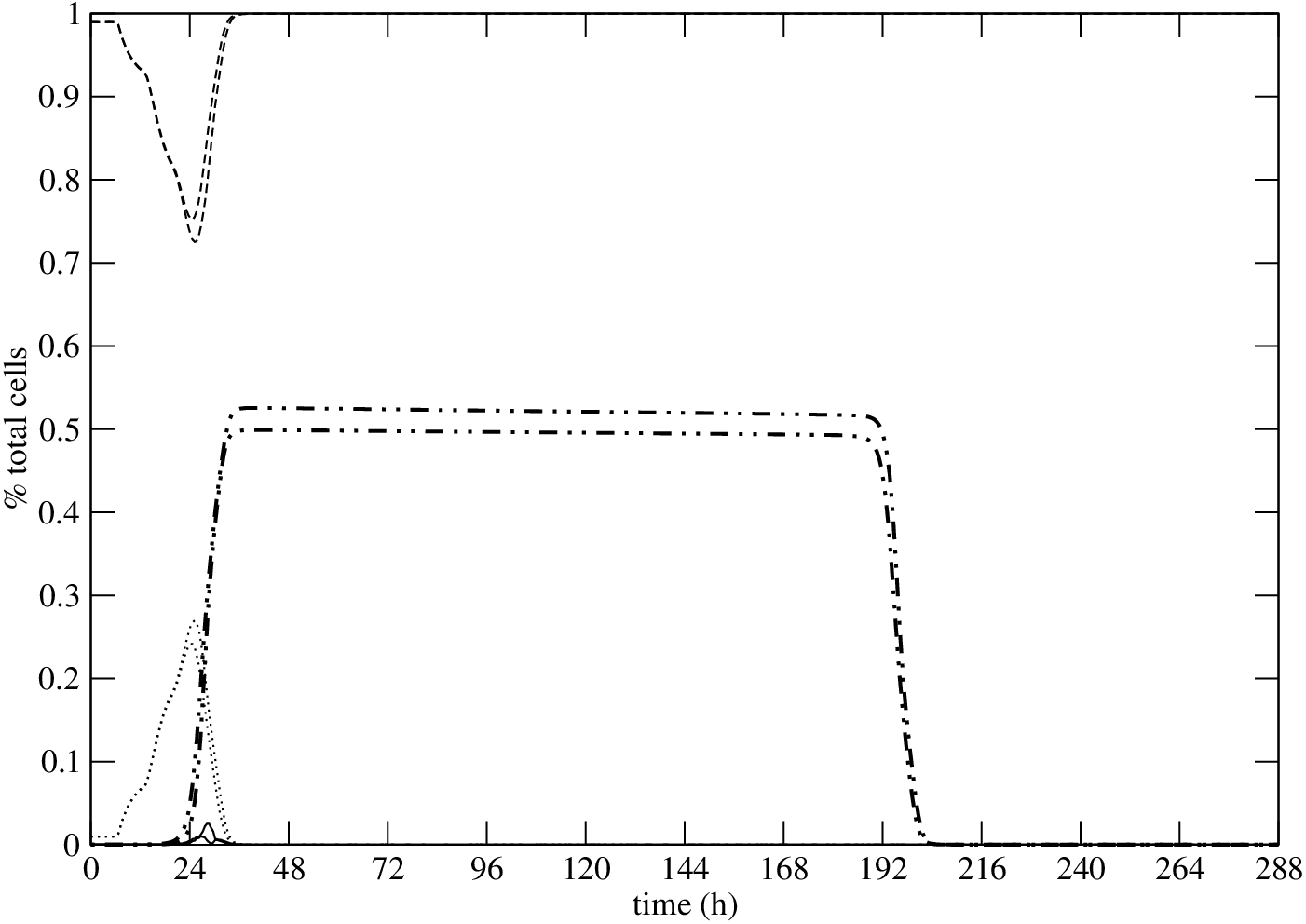}}
\end{center}
\caption{The effect of varying \texttt{RECRUITMENT} on the viral infection's dynamics. From left to right, the graphs show the behaviour obtained using \texttt{RECRUITMENT} values of $0.05$, $0.25$, and $1.25$ respectively. The central graph is the same as Figure \ref{simulation}. These show that the simulation is sensitive to the parameter \texttt{RECRUITMENT}.}
\label{g_rec}
\end{figure}

The typical immune cell division time is about $\unit{12}{\hour} \sim \unit{24}{\hour}$ \cite{bocharov94}, which corresponds to a recruitment ranging from 0.08 to 0.16 immune cells per hour. Considering that an immune cell encounters 6 epithelial cells per hour ($\texttt{FLOW\_RATE} = 6$ time steps per hour) and that on average about $15$\% of those cells are infected, then our model's effective recruitment rate is about $0.25 \times 6 \times 0.15 = 0.23$ immune cells per hour, which is higher than the biological range described above. However, that recruitment range is based soly on the cellular divisions of activated B and T cells and does not take into consideration the high production of antibodies by plasma B cells which contributes to reducing the infection, a factor that can account for the higher recruitment rate value required by our model. Simulation runs using a \texttt{RECRUITMENT} value of $0.05$ and $1.25$ are presented in Figure \ref{g_rec}.
One can see that the parameter \texttt{RECRUITMENT} has an important effect on the development and the outcome of the infection: too small a value makes the infection fatal, while larger values decrease the amplitude and duration of the infection. It is thus a key parameter and much effort should be invested in finding a way to draw a parallel between it and true biological parameters.

\subsection{\texttt{RECRUIT\_DELAY}}
\begin{figure}
\begin{center}
\resizebox{0.3\textwidth}{!}{\includegraphics{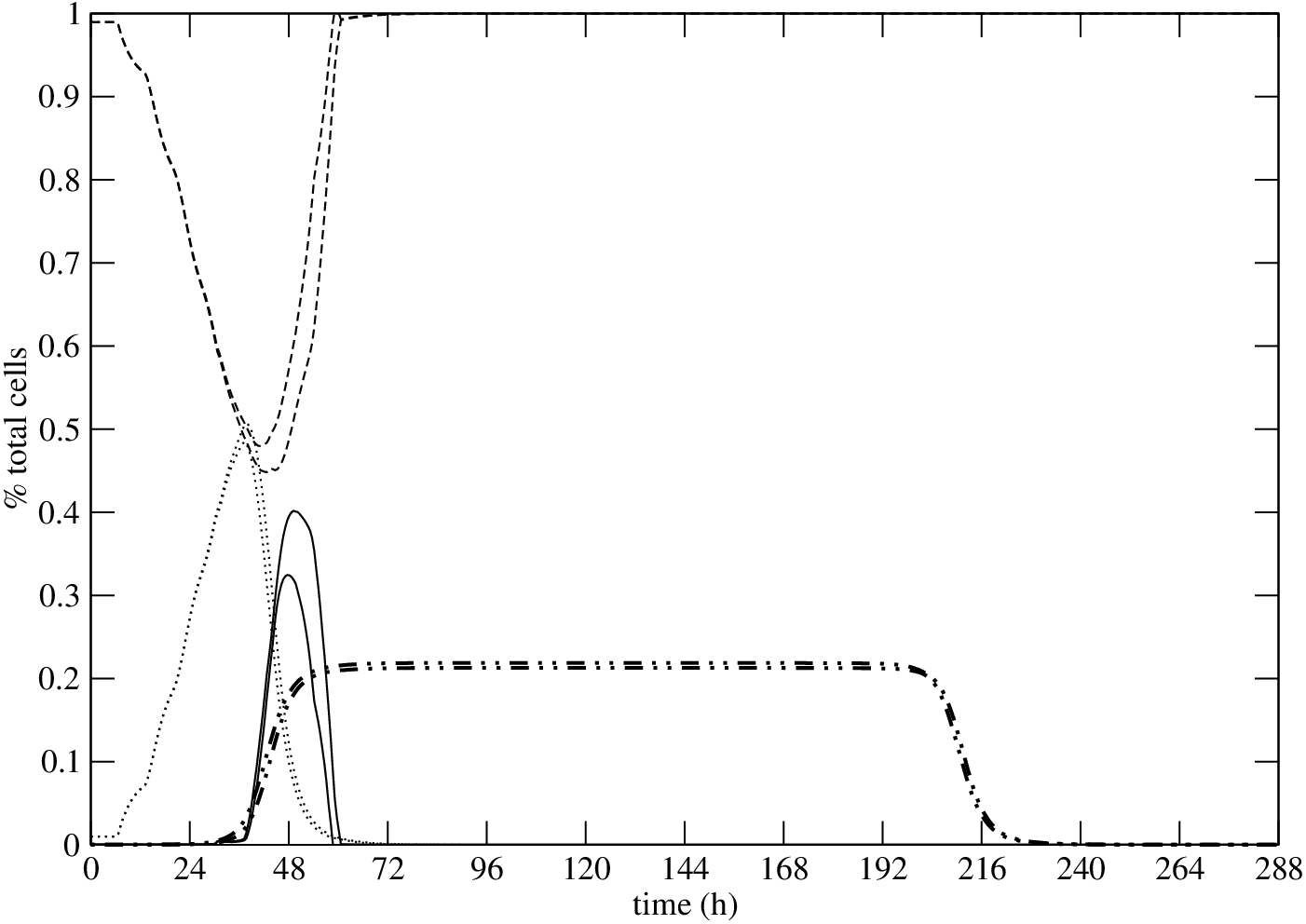}}
\resizebox{0.3\textwidth}{!}{\includegraphics{simulation}}
\resizebox{0.3\textwidth}{!}{\includegraphics{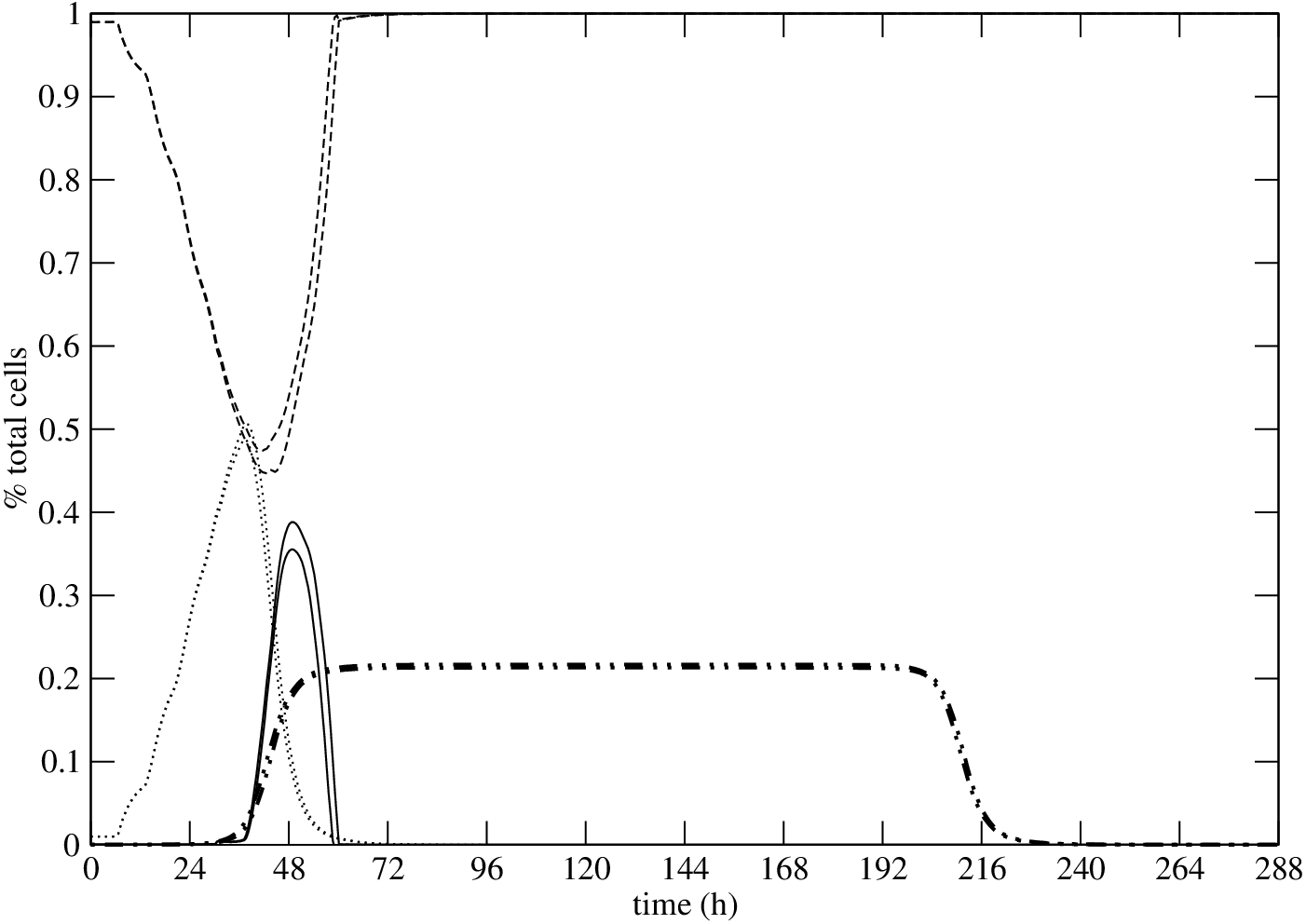}}
\end{center}
\caption{The effect of varying \texttt{RECRUIT\_DELAY} on the viral infection's dynamics. From left to right, the graphs show the behaviour obtained using \texttt{RECRUIT\_DELAY} values of \unit{2}{\hour}, \unit{7}{\hour}, and \unit{12}{\hour} respectively. The central graph is the same as Figure \ref{simulation}. These show that the simulation is not sensitive to the parameter \texttt{RECRUIT\_DELAY}.}
\label{g_del}
\end{figure}

Once a ``recruitment call'' has been made, immune cells are not added immediately to the simulation lattice. The delay between the call and the actual addition of immune cells corresponds roughly to the typical transfer time of cytotoxic T lymphocytes from their main area of production --- lung associated lymphoid tissue (LALT) compartment --- to the affected area --- the mucous compartment --- which, in \cite{bocharov94}, is said to take from \unit{2}{\hour} to \unit{12}{\hour}. We chose to use the mid-range value of \unit{7}{\hour} \texttt{RECRUIT\_DELAY}'s default value. Figure \ref{g_del} shows the effect of varying \texttt{RECRUIT\_DELAY} over the biologically acceptable range.
It is clear that changes to the value of this parameter have little influence on the dynamics of the model over the uncertainty range.

\section{Conclusion}
Here, we have introduced a CA model for an uncomplicated viral infection. We have shown that once parameterized for the particular case of influenza A, our CA model --- which is described by 7 state variables and 12 parameters --- is sophisticated enough to reproduce the basic dynamical features of the infection. One expects a 12 parameter model to be able to match 7 dynamical features, however all but 5 parameters of this model are sufficiently-well bound by physiological data that they cannot be used to tune its behaviour. Remarkably, our 7 variable and 12 parameter model's agreement with the experimental dynamical characteristics compares well to that of the 13 variable and 60 parameter model presented in \cite{bocharov94}. To the authors' knowledge, the CA model presented here along with the ODE model presented in \cite{bocharov94} are the only two existing immunological models for influenza A viral infections. This is quite surprising considering its significance to human health worldwide, particularly its high potential for causing devastating epidemics \cite{who_flu}.

The results obtained with the model described herein have surpassed our expectations. We had hoped to develop a model that would be good enough to be used as a test bench to investigate various theoretical aspects of viral infections, and we have obtained an influenza A model that behaves impressively well when compared against available experimental data. We believe that by adding additional details to the model such as specific immune cell types, our CA model could be a very promising model of influenza A.

A wide range of behaviours were observed as the model's parameters were varied over their biologically allowed range. It would be valuable to test whether these predicted behaviours can actually be observed \latin{in vivo}, since if they are not seen then that would falsify the model.

In future work, we will make use of the CA model introduced here to investigate various theoretical aspects of viral infections, for example the extent to which such things as spatial inhomogeneities and age classes can affect the evolution and outcome of a viral infection. It will be interesting to compare this CA model against the recent work in \cite{chao03} which makes use of a stochastic stage-structured model, an alternative to both differential equations and cellular automata.

\begin{acknowledgments}
This work was supported in part by MITACS' Mathematical Modelling in Pharmaceutical Development (MMPD) project. The first author would also like to thank Dr.\ Kipp Cannon for helpful discussions.
\end{acknowledgments}

\addcontentsline{toc}{section}{References}
\bibliographystyle{abbrv}
\bibliography{influenzaA}

\end{document}